\documentclass{aa}
\usepackage{savesym}
\usepackage{mathtools}
\usepackage{amssymb}
\savesymbol{iint}
\usepackage{graphicx}
\usepackage{txfonts}
\restoresymbol{TXF}{iint}
\begin{document} 
\title{Dust cloud evolution in sub-stellar atmospheres via plasma deposition and plasma sputtering}
   \author{C. R. Stark
          \inst{1},
          \and
          D. A. Diver\inst{2}
          }
   \institute{Division of Computing and Mathematics, Abertay University, Kydd Building, Dundee DD1 1HG.\\
              \email{c.stark@abertay.ac.uk}
         \and
             SUPA, School of Physics and Astronomy, Kelvin Building, University of Glasgow, Glasgow, G12 8QQ, Scotland, UK.\\ 
\email{declan.diver@glasgow.ac.uk}             
             }
   \date{}
  \abstract
  % context heading (optional)
  % {} leave it empty if necessary  
   {In contemporary sub-stellar model atmospheres, dust growth occurs through neutral gas-phase surface chemistry. Recently, there has been a growing body of theoretical and observational evidence suggesting that ionisation processes can also occur. As a result, atmospheres are populated by regions composed of plasma, gas and dust, and the consequent influence of plasma processes on dust evolution is enhanced.}
  % aims heading (mandatory)
   {This paper aims to introduce a new model of dust growth and destruction in sub-stellar atmospheres via plasma deposition and plasma sputtering.}
  % methods heading (mandatory)
   {Using example sub-stellar atmospheres from \textsc{Drift-Phoenix}, we have compared plasma deposition and sputtering timescales to those from neutral gas-phase surface chemistry to ascertain their regimes of influence. We calculated the plasma sputtering yield and discuss the circumstances where plasma sputtering dominates over deposition.}
  % results heading (mandatory)
   {Within the highest dust density cloud regions, plasma deposition and sputtering dominates over neutral gas-phase surface chemistry if the degree of ionisation is $\gtrsim10^{-4}$.  Loosely bound grains with surface binding energies of the order of $0.1-1$~eV are susceptible to destruction through plasma sputtering for feasible degrees of ionisation and electron temperatures; whereas, strong crystalline grains with binding energies of the order $10$~eV are resistant to sputtering.}
  % conclusions heading (optional), leave it empty if necessary 
   {The mathematical framework outlined sets the foundation for the inclusion of plasma deposition and plasma sputtering in global dust cloud formation models of sub-stellar atmospheres.}

    \titlerunning{plasma sputtering and plasma deposition}
 \maketitle

%
%________________________________________________________________

\section{Introduction}
Understanding the formation, growth and destruction of dust, leading to the evolution of large-scale cloud structures in sub-stellar atmospheres is key to interpreting their electromagnetic spectra and characterising their role in the transition between L and T dwarfs.  sub-stellar dust formation models assume dust nucleation, mantle growth and evaporation occurs in the gas-phase~\citep{helling2004,dehn2007,helling2006,helling2008b,helling2008c,witte2009,witte2011,tsuji2002,allard2001,burrows1997,marley2002,morley2012}; however, ionisation processes may occur that produce regions of atmospheric plasma \citep{helling2011a,helling2011b,helling2013,bailey2013,rimmer2013,stark2013,isabel2015}. The presence of an ionised component in sub-stellar atmospheres is backed-up by observations of radio and X-ray signatures for example \citet{williams2015}. Dust immersed in a plasma will become negatively charged and its subsequent growth will be affected by the plasma environment.  Depending on the degree of ionisation, plasma processes such as plasma deposition (and not classical gas-phase processes) may be the dominant dust nucleation and growth processes.  

A number of ionisation processes can occur in sub-stellar atmospheres to generate plasma regions including thermal ionisation~\citep{isabel2015}, gas discharge events (lightning)~\citep{helling2013}, cosmic-ray ionisation~\citep{rimmer2013}, turbulence-induced dust-dust collisions~\citep{helling2011b} and Alfv\'{e}n ionisation~\citep{stark2013}. Photoionisation from a companion or host star is an important aspect of all ionisation processes, particularly lightning, since it helps to create a seed population of electrons required to kickstart further ionisation. Thermal ionisation typically occurs deep in the atmosphere ($p_{\rm gas}\approx10^{-4} - 10$~bar) and can create regions of plasma whose spatial extent and degree of ionisation depends upon the effective temperature, metallicity and $\log{g}$ of the object. For example, increasing the effective temperature from $T_{\rm eff}\approx 1200$ to $3000$~K, the atmospheric volume fraction that exhibits plasma behaviour (long-range, collective effects) increases from $10^{-3}-1$~\citep{isabel2015}. At best, the thermal degree of ionisation can reach values of $10^{-7}-10^{-4}$. Within cloud regions ($p_{\rm gas}\approx10^{-5}-1$~bar), lightning also contributes to the overall ionisation fraction of the atmosphere yielding degrees of ionisation of about $10^{-1}$~\citep{guo2009,beyer2003}, where a single discharge event can have characteristic vertical length scales $\approx 0.5 - 4$~km, affecting atmospheric volumes of the order of $10^{4}-10^{10}$~m$^{3}$ \citep{bailey2013}. For multiple discharge events, the atmospheric volume affected can be significantly enhanced; for example, \citet{gabi2016} estimate the total number of lightning flashes of exoplanets to be of the order of $10^{5}-10^{12}$ during a transit. Alfv\'{e}n ionisation (AI) occurs when a neutral gas collides with a low-density magnetised seed plasma and their relative motion reaches a critical threshold speed. It can produce regions with degrees of ionisation ranging from $f_{e}\approx10^{-6}-1$ and is most effective where $p_{\rm gas}\approx10^{-5}-10^{-15}$~bar and where flow speeds are sufficiently high, $\approx\mathcal{O}(1-10$~km$^{-1})$~\citep{stark2013}. The distribution and spatial extent of plasma regions generated by thermal ionisation, lightning and AI will depend upon the temperature distribution, cloud distribution and the wind flow profiles in the atmosphere respectively. For lightning, regions of dense cloud coverage (e.g. see Figure 6,~\citet{Lee2016}) may encourage higher lightning rates and hence indicate plasma region hotspots.  Similarly for thermal ionisation, regions of higher temperature (e.g. see Figure 3,~\citet{kaspi2015} and Figure 2,~\citet{Lee2016}) may have a greater degree of ionisation; and for AI, flow speeds $\approx1-10$~kms$^{-1}$ (e.g. see Figure 3,~\citet{dobbs2012}) will yield degrees of ionisation of the order of $10^{-4}$.

Dust growth in plasmas occurs through a series of distinct stages: an initial growth phase up to a critical dust number density, where seed particles form through plasma chemistry driven nucleation; a rapid growth phase through the coagulation of the small-sized population;  and then a fast growth phase via plasma deposition resulting in macroparticles of micrometre size~\citep{watanabe2006,bingham2001,sorokin2004,hollenstein2000,tsytovich_2004_rev,bouchoule1999}.  The coagulation phase is restricted by the dust acquiring a net charge, after which plasma deposition then becomes the dominant process producing micrometre-sized dust grains.  In the plasma deposition stage, ions from the bulk plasma are accelerated through the dust grain's plasma sheath and ultimately deposited on the grain surface, forming a layer of material (e.g.~\citet{cao2002,cao2003,matsoukas2004}). The local structure of the electric field at the grain's surface can affect the distribution of deposited material, producing non-spherical geometries~\citep{Stark2006}. In low-pressure laboratory conditions ultrathin films of the order of a nanometre have been deposited on the surfaces of nanoparticles via plasma polymerisation; for example, pyrrole coated alumina nanoparticles~\citep{shi2001}; and copper particles coated in C$_{2}$H$_{6}$ monomer layers~\citep{qin2007}.  

Dust destruction~\citep{salpeter1977,draine1979b,dwek1992} can take many forms: thermal evaporation (e.g.~\cite{fleischer1992,rapacioli2006}); chemical sputtering  (e.g.~\cite{bauer1997,nanni2013}); physical sputtering (e.g.~\cite{barlow1978,bringa2002,silvia2012}); grain-grain collisions (e.g.~\cite{draine1979,tielens1994,deckers2014}); and shock disruption (e.g.~\cite{jones1994,jones1996,raymond2013}).  In plasma-based material processing technology, plasma sputtering or etching is the process where atoms are removed from the surface of a material using energetic ions (e.g.~\cite{ohring,jacob1998,chu2002,mattox,wasa,seshan,donnelly2013}) such as Sulphur hexafluoride (SF$_{6}$) or Argon to etch features in silicon substrates. In a conventional sputtering system, plasma ions are accelerated towards the target material thanks to a high negative potential (see~\cite{chapman1980,sugawara1998}). If the energy of the ions is high enough they will sputter the target atoms. In such plasma processing ion beam energies of $\approx\mathcal{O}(10$~eV) results in single knock-on sputtering; for energies $\approx500-1000$~eV, the incident ion can excite a large group of surface atoms and molecules, allowing a larger number of atoms to be liberated~\citep{fridman2008}.  These energies are relevant for surfaces used in the nanofabrication of semiconductors where lattice energies are of the order $E_{\rm latt}\approx10$~eV; however, for natural dust, energies may be lower.  

The aim of this paper is to investigate dust growth and destruction via plasma deposition and plasma sputtering in sub-stellar atmospheric plasmas.  This paper introduces two established laboratory plasma processes (plasma deposition and plasma sputtering) into the brown dwarf context, where it has not been considered before.  Plasma assisted deposition and sputtering is known to be much more effective than conventional gas-phase chemistry, since it is free of the thermodynamic restrictions of the latter, and able to harness electrostatic fields to produce higher-energy particle impacts, combined with electron-moderated chemistry (e.g. see~\cite{stoffel1996,alexandrov2005,jones2009}).  In contrast to previous studies, this paper considers how plasma collective effects self-consistently energise the plasma ions such that they can participate in plasma deposition and plasma sputtering.  In Section~\ref{sec_theo} a theory of plasma deposition and sputtering is outlined.  In Section~\ref{sec_res}, plasma deposition and sputtering is discussed in the context of example sub-stellar atmospheres from the \textsc{Drift-Phoenix} model atmosphere and cloud formation code.  

\section{Plasma deposition and plasma sputtering\label{sec_theo}}

In the gas-phase, grain growth occurs via the absorption of neutral species via the stochastic kinetic motions of the gas.  However, in an atmospheric dusty gas-plasma region (a plasma containing dust particles), the absorption of species can be electrostatically driven due to the grains gaining a net negative charge. Dust immersed in an electron-ion plasma naturally acquires a net negative charge due to the greater mobility of the electrons relative to the ions.  For a given thermodynamic temperature, the electrons have a greater probability of striking and sticking to the grain surface.  However, as the dust grain accumulates more and more electrons, its negative charge increases and the probability of further electron attachment decreases.  Meanwhile, the plasma ions are accelerated towards the negatively charged dust grain and are deposited on the grain's surface.  In doing so the ions reduce the net negative charge on the dust grain, inadvertently increasing the probability of further electron attachment and leading to the continual adjustment of the grain's charge.  The net charge of the dust grain continues to fluctuate until a particle-flux equilibrium configuration is reached where the electron and ion fluxes at the dust grain surface are equal.  At this stage, the dust grain has a constant negative charge and the grain's surface is at the floating potential~\citep{lieb2005},
\begin{equation}
\phi_{f}=-\frac{k_{B}T_{e}}{2e}\ln{\left(\frac{m_{i}}{2\pi m_{e}}\right)},
\end{equation}  
where $k_{B}$ is the Boltzmann constant; $T_{e}$, the electron temperature; $m_{i,e}$, the ionic and electron mass respectively; and $e$, the charge of an electron. The dust grain is now surrounded by a plasma sheath (an electron depleted region) with a sheath length, $d_{\rm sh}$, of the order of the plasma Debye length, $\lambda_{D}$, such that on length scales greater than the sheath length the dust grain is screened by the bulk plasma. The Debye length is an e-folding distance defined as $\lambda_{D}=(\epsilon_{0}k_{B}T_{e}/(n_{e}e^{2}))^{1/2}$, where $\epsilon_{0}$ is the permittivity of free space and $n_{e}$ is the electron number density.  
We note that in this context screening is the damping of electric fields by the free charges in the plasma. The plasma arranges itself in such a way that the electric field of the charged dust grain is very small, but not negligible, on scales greater than the Debye length away from the grain. The residual undamped component of the electric field accelerates the ions from the bulk plasma into the sheath region. In comparison to the neutral, gas-phase case, the ionic flux is enhanced and the ionic energy amplified, increasing the absorption of species to the grain surface and increasing the growth rate of the dust grain.  Furthermore, the probability of surface chemical reactions increases and chemical reactions with greater activation energies (that would otherwise be inaccessible thermally) are more likely~\citep{stark2014}.  

The edge of the plasma sheath occurs where the thermal energy, $k_{B}T_{e}/2$, of the electrons is comparable to their electric potential energy.  At this boundary, the residual electric field from the charged dust grain is weak, and so a fraction of electrons have enough thermal energy to surmount the electrostatic potential.  Therefore, electric potentials of the order of $k_{B}T_{e}/(2e)$ are not shielded and can leak into the plasma.   In the case of the shielding of a charged dust particle, this potential is responsible for initially accelerating the ions from the plasma and into the sheath towards the particle surface.  As a result of this potential, $\phi$, the change in speed of the ions upon entering the sheath is given by energy conservation as $q_{i}\phi=k_{B}T_{e}/2=m_{i}v^{2}/2$, where $q_{i}$ is the charge on an ion and $m_{i}$ is the mass of an ion.  Therefore, the change in speed of the plasma ions entering the plasma sheath is $u_{0}=(k_{B}T_{e}/m_{i})^{1/2}$, the Bohm speed.  The ions from the plasma are accelerated towards the grain and interact with the grain surface altering the shape and size of the grain.  If the energy of the ions is below the surface binding energy of the atoms on the grain surface the incoming ion will be deposited on the surface, growing a layer of material; if the ionic energy exceeds the surface binding energy then atoms will be sputtered (ejected) from the surface.

Consider the flux of ions with number density $n_{i}$ and average speed $\langle v\rangle$, $F=\frac{1}{4}n_{i}\langle v\rangle$, incident on the surface of a dust grain of radius $a$~\citep{woods1993,chen1984}.  Assuming that all ions incident on the grain surface must interact with it, the reaction rate is 
\begin{equation}
R=\pi a^{2}n_{i}\langle v \rangle. \label{tot_reac_0}
\end{equation}
The total reaction rate can be expressed in terms of the reaction rate for ionic deposition on the surface, $R_{\rm dp}$ and for the sputtering of atoms and molecules from the grain surface due to ion bombardment, $R_{\rm sp}$,
\begin{eqnarray}
R&=&R_{\rm dp}+R_{\rm sp} \\
&=&\langle \sigma_{\rm dp}v\rangle n_{i}+\langle \sigma_{\rm sp}v\rangle n_{i}, \label{tot_reac}
\end{eqnarray}
where $\sigma_{\rm dp}$ and $\sigma_{\rm sp}$ are the reaction cross-sections for plasma deposition and sputtering from the grain surface respectively.  The reaction cross-section is the area perpendicular to the relative motion of the two particles that they must occupy in order to interact. Here, the cross-sections are dependent on the incoming ion energy and the threshold energy required for the interaction which is related to the binding energy of the target species. 

The resulting contribution to the total mass of the dust grain $M$ with time $t$ due to deposition and sputtering can be calculated,
\begin{eqnarray}
\frac{\textnormal{d}M}{\textnormal{d}t}&=&m_{i}R_{\rm dp}-m_{\rm n}Y_{\rm sp} R_{\rm sp} \\
&=&n_{i}(m_{i}\langle \sigma_{\rm dp}v\rangle-m_{\rm n}Y_{\rm sp}\langle \sigma_{\rm sp}v\rangle), \label{sub_react}
\end{eqnarray}
where $m_{\rm n}$ and $m_{i}$ is the mass of the sputtered atom or molecule and the deposited ion respectively.  This approach is equivalent to calculating the number of beam particles passing through a target area per unit time and multiplying it by the probability of an interaction (the sticking coefficient) to get the number of interactions per unit time. The parameter $Y_{\rm sp}$ is the sputtering yield and is defined as the number of target atoms (or molecules) ejected per incident ion. In a sputtering event, the incoming ion either remains on the surface or recoils into the surrounding medium. So, in some interactions the mass of the grain will not change. The competition between deposition and sputtering determines the net effect on the grain mass. Equation (\ref{sub_react}) encapsulates this since the reaction rates are energy dependent.

For low-energy ions of the order of the threshold energy for sputtering, $E_{\rm th}$, the sputtering yield is given by~\citep{tielens1994},
\begin{eqnarray}
Y_{\rm sp}&\approx&\frac{1.12}{E_{\rm b}}\frac{x^{2}}{(x+1)(x^{2/3}+1)^{1/2}}\left(\frac{m_{n}}{u}\right)^{5/3}\alpha\frac{R_{\rm p}}{R}s(\epsilon)\nonumber\\
&&\times\left[1-\left(\frac{E_{\rm th}}{E_{\rm sf}}\right)^{2/3}\right]\left(1-\frac{E_{\rm th}}{E_{\rm sf}}\right)^{2},~~~~~~~~~~~~~~~~E>E_{\rm th} \label{sputt_eqn}
\end{eqnarray}
where $x=m_{\rm i}/m_{\rm n}$, and
\begin{eqnarray}
\frac{R_{\rm p}}{R}&=&(K/x+1)^{-1} \\
s(\epsilon)&=&\frac{3.441\sqrt{\epsilon}\ln{(\epsilon+2.718)}}{1+6.35\sqrt{\epsilon}+\epsilon(-1.708+6.882\sqrt{\epsilon})} \\
\epsilon&\approx&\frac{4}{(x+1)x}\left(\frac{u}{m_{n}}\right)^{2}\frac{4\pi\epsilon_{0}a_{\rm SL}}{e}E_{\rm sf}\\
a_{\rm SL}&\approx&1.115a_{0}(1+x^{2/3})^{-1/2}\left(\frac{m_{n}}{u}\right)^{-1/3} \\
E_{\rm th}&=&\begin{dcases} \dfrac{E_{\rm b}}{g(1-g)},~~~~~~x\leq0.3 \\8E_{\rm b}x^{1/3},~~~~~~x>0.3 \end{dcases} \\
g&=&\frac{4x}{(1+x)^{2}}\\
\alpha&=&\begin{cases}0.3x^{-2/3},~~~~~~~2>x>0.1 \\ 0.2,~~~~~~x>2 \end{cases}
\end{eqnarray}
where $a_{0}=0.529\times10^{-10}~\rm{m}$ is the Bohr radius; $u$ is the atomic mass unit; $E_{\rm sf}$ is the energy of the incident ion; $E_{\rm b}$ is the surface binding energy of the material being sputtered; $E_{\rm th}$ is the threshold energy required for sputtering; 
$R_{\rm p}/R$ is the ratio of the mean projected range to the mean penetrated path length; $a_{\rm SL}$ is the screening length for the interaction potential between the nuclei; and, $K=0.4$. All energies are in electronvolts, eV. In the sputtering expressions the atomic mass number of the participating species $Z_{s}$, has been approximated by $Z_{s}\approx m_{s}/(2u)$.  For further details on sputtering see also~\cite{wasa,seshan,ohring,mattox,barlow1978,sigmund1969,chapman1980,harsha}.

Assuming a spherical grain with radius $a$ and constant material mass density $\rho$, the mass of the dust grain is $M=\frac{4}{3}\pi\rho a^{3}$; therefore,
\begin{eqnarray}
\frac{\textnormal{d}}{\textnormal{d}t}\left(\frac{4}{3}\pi\rho a^{3}\right)&=&4\pi a^{2}\rho\frac{\textnormal{d}a}{\textnormal{d}t}\\
&=&n_{i}(m_{i}\langle \sigma_{\rm dp}v\rangle -m_{\rm n}Y_{\rm sp}\langle \sigma_{\rm sp}v\rangle).
\end{eqnarray}
Equations~(\ref{tot_reac_0})~and~(\ref{tot_reac})~give
\begin{equation}
\pi a^{2}=\frac{\langle \sigma_{\rm dp} v\rangle+\langle \sigma_{\rm sp} v\rangle}{\langle v\rangle}.
\end{equation}
As a result, the temporal evolution of the grain radius $a$ with time $t$ is given by
\begin{equation}
\frac{\textnormal{d}a}{\textnormal{d}t}=\frac{n_{i}\langle v\rangle}{4\rho}\left(\frac{m_{i}\langle\sigma_{\rm dp}v\rangle-m_{\rm n}Y_{\rm sp}\langle\sigma_{\rm sp}v\rangle}{\langle\sigma_{\rm dp}v\rangle+\langle\sigma_{\rm sp}v\rangle}\right).\label{adot}
\end{equation}
The energy of the ions upon reaching the surface of the dust grain ($E_{\rm sf}$) is given by,
\begin{equation}
E_{\rm sf}\approx (1-p)q_{i}|\phi_{f}|, \label{e_surf}
\end{equation}
where $q_{i}$ is the electric charge of the ion (here it is assumed that $q_{i}=e$) and $p$ is the probability of the incoming ion colliding with a neutral atom in the sheath,
\begin{equation}
p=\frac{\delta_{\rm ni}}{1+\delta_{\rm ni}}.
\end{equation}
$\delta_{\rm ni}=d_{\rm sh}/\lambda_{\rm mfp,n}$ is the ratio of the plasma sheath length and the collisional lengthscale for the ions with neutral atoms, $\lambda_{\rm mfp,n}$.  The factor $p$ incorporates the effect of collisions between ions and neutrals in the sheath. The more collisions that an ion experiences while traversing the sheath, the less likely it will be able to attain the energy it would if it was accelerated unimpeded through the sheath to the grain surface.  The effect of collisions on the ion motion is more significant when the degree of ionisation is low.  In the calculations presented here, the approximations $d_{\rm sh}\approx\lambda_{D}$ and $\lambda_{\rm mfp,n}\approx(n_{\rm n}\sigma_{\rm n})^{-1}$ are made, where $\sigma_{\rm n}\approx\pi r_{\rm n}^{2}$ and $r_{\rm n}\approx10^{-10}$~m is the radius of a neutral atom. The gas number density is given by $n_{\rm gas}=n_{\rm n}+2n_{e}$ and the degree of ionisation can be written as $f_{e}=n_{e}/(n_{\rm n}+n_{e})$.  Therefore,   
\begin{equation}
\delta_{\rm ni}\approx10^{-18}\frac{\left(1-f_{e}\right)}{f_{e}^{1/2}\left(1+f_{e}\right)^{1/2}}\left(n_{\rm gas}T_{e}\right)^{1/2},\label{delta_ni}
\end{equation}
where $n_{\rm gas}$ is in units of [m$^{-3}$].  At the surface of the dust grain there will be a distribution of ion energies. If the energy imparted to the surface atoms and molecules exceeds the required threshold energy they will overcome the forces binding them and they will be sputtered.  Depending on the threshold energy required to sputter material off the grain's surface, there will be a fraction of ions that will be deposited on the grain's surface and a fraction that will sputter material.  If the mean energy of the ions is smaller than the threshold for sputtering, the majority of ions will be deposited on the grain's surface and only a small population will have sufficient energy to sputter atoms.  If the mean energy of the ions is greater than the threshold for sputtering, the majority of ions will have sufficient energy to sputter atoms off the surface and only a small fraction will be deposited. 

In the regime where the average kinetic energy of the ions incident on the grain surface is less than the surface binding energy of the surface atoms and molecules ($E_{\rm sf}\ll E_{\rm th}$), Eq.~(\ref{adot}) becomes
\begin{equation}
\frac{\textnormal{d}a}{\textnormal{d}t}=\gamma_{\rm dp}=\frac{m_{i}n_{i} \langle v\rangle}{4\rho}. \label{dep}
\end{equation}

When the average kinetic energy of the ions exceeds the surface binding energy of surface atoms and molecules ($E_{\rm sf}\gg E_{\rm th}$), Eq.~\ref{adot} simplifies to
\begin{equation}
\frac{\textnormal{d}a}{\textnormal{d}t}=\gamma_{\rm sp}=-\frac{m_{\rm n}n_{i} \langle v\rangle}{4\rho}Y_{\rm sp}. \label{sputt}
\end{equation}
We note that $n_{i}\langle v\rangle=n_{i0}u_{0}=$~constant, where $n_{i0}$ is the ion number density of the bulk plasma ($n_{i0}\approx n_{e0}$ for quasineutrality in a plasma); and $u_{0}=(k_{B}T_{e}/m_{i})^{1/2}$ is the Bohm speed of the ions upon entering the sheath. Therefore, the growth or sputtering rate becomes,
\begin{equation}
\gamma_{\rm dp,sp}=\frac{m_{\rm i,n}n_{i0}u_{0}}{4\rho}, \label{rates}
\end{equation}
for maximal values, when $Y_{\rm sp}\approx1$. 

In \textsc{Drift-Phoenix} model sub-stellar atmospheres the mass density of a dust grain is of the order of $10^{3}$~kg~m$^{-3}$ and $m_{\rm i,n}\approx\mathcal{O}(10^{-26}$~kg). The chemical make-up of the dust grains has little effect on plasma deposition: a mantle will grow irrespective of surface chemical composition. For plasma sputtering the bonding of the surface (hence the chemical composition) will determine the ionic energy needed to break the bonds required to eject material. However, in this paper a wide range of bonding energies are considered, taking into account the chemical diversity of the dust. Furthermore, as ionic material is deposited on the dust surface, the chemical composition of the grown mantle will depend on the ionic species being deposited and so the initial chemical composition of the grain has no further effect on the process.

The key physical parameters that determine the deposition and sputtering rates are the ion number density $n_{i0}$ and the electron temperature $T_{e}$.  The ion number density can be written in terms of the degree of ionisation, $f_{e}=n_{e}/(n_{\rm gas}-n_{e})$ and so the maximum time taken (where $Y_{\rm sp}\approx1$) for plasma deposition or sputtering to grow or remove a monolayer (width~$\approx10^{-10}$~m) of material is
\begin{equation}
\tau_{\rm mono}=\frac{10^{-10}}{\gamma_{\rm sp,dp}}\approx 10^{17}\frac{(1+f_{e})}{f_{e}}\frac{1}{n_{\rm gas}T_{e}^{1/2}}~~~[s], \label{eq_mono}
\end{equation}
where $n_{\rm gas}$ is in units of [m$^{-3}$]; and, we have made the simplification $m_{\rm i}\approx m_{\rm n}\approx\mathcal{O}(10^{-26})$~kg.  A measure of the relative strength of plasma sputtering relative to plasma deposition can be given by,
 \begin{equation}
 \Gamma=\frac{m_{\rm n}}{m_{\rm i}}\frac{R_{\rm sp}}{R_{\rm dp}}Y_{\rm sp}.
 \end{equation}
 This expression can be simplified by setting $R_{\rm sp}/R_{\rm dp}=1$; therefore, 
\begin{equation}
\Gamma\approx\frac{m_{\rm n}}{m_{\rm i}}Y_{\rm sp}=\frac{Y_{\rm sp}}{x}. \label{big_g}
\end{equation}
For the sputtering regime, $\Gamma>1$; and the deposition regime, $\Gamma<1$. The case $\Gamma=1$ presents the equilibrium solution of Equation~\ref{adot} ($\dot{a}=0$) where the radius of the dust grain neither increases nor decreases in size.

\section{Results \label{sec_res}}
\begin{figure}
\resizebox{\hsize}{!}{\includegraphics{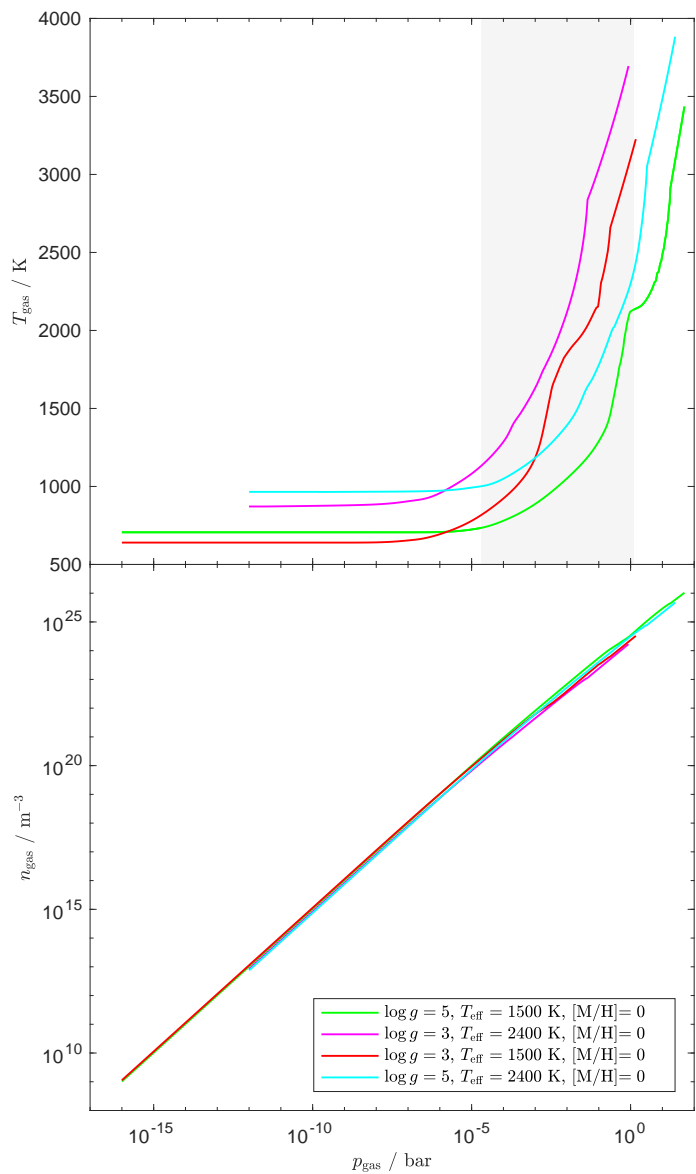}}
\caption{Pressure-temperature diagram for example sub-stellar atmospheres obtained using \textsc{Drift-Phoenix} model atmosphere and cloud formation code; Bottom: The gas-phase number density as a function of atmospheric pressure corresponding to the pressure-temperature diagrams for the example sub-stellar atmospheres plotted in the top plot. The grey patch shows the region where the highest density cloud layers are present for the example atmospheres.\label{dep_sputt_0}}
\end{figure}

In this paper we are interested in plasma deposition and sputtering in sub-stellar atmospheres. We consider example atmospheres (Figure~\ref{dep_sputt_0}) using the \textsc{Drift-Phoenix} model atmosphere and cloud formation code~\citep{hauschildt1999,helling2004,helling2008d,helling2006,dehn2007,witte2009,witte2011} characterised by $T_{\rm eff}=1500$~K and $T_{\rm eff}=2400$~K, for both $\log{g}=3.0$ and $\log{g}=5.0$; and solar metallicities ([M/H]=0.0). These example atmospheres cover a wide and inclusive range of parameter space.

\begin{figure}
\resizebox{\hsize}{!}{\includegraphics{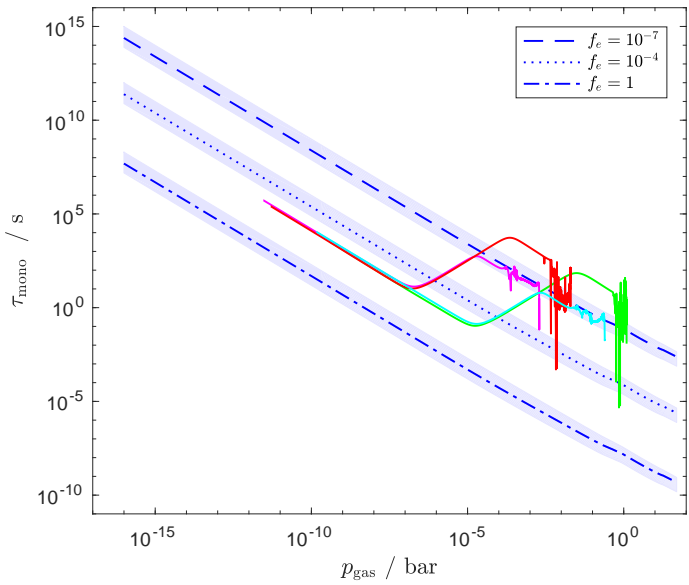}}
\caption{Time taken for plasma deposition (sputtering) to grow (remove) a monolayer, $\tau_{\rm mono}$~(Equation~\ref{eq_mono}), as a function of atmospheric gas pressure, $p_{\rm gas}$. We note that $p_{\rm gas}$ acts as a proxy for atmospheric height and in general does not reflect the true local pressure. $\tau_{\rm mono}$ is shown for an electron temperature $T_{e}=1$~eV and degrees of ionisation $f_{e}=10^{-7}$ (dash), $10^{-4}$ (dot) and 1 (dot-dash). The light blue bands shows the effect of varying the electron temperature from $T_{e}=T_{\rm gas}$ (upper boundary) to $T_{e}=10$~eV (lower boundary), for each degree of ionisation.  In this paper $f_{e}$ is treated as a parameter and is varied, from the values calculated in \textsc{Drift-Phoenix}, to incorporate expected values from various non-thermal ionisation processes beyond thermal ionisation. As a consequence, as the number density of the electrons, ions and neutral particles change, the local gas pressure will change and deviate from its original \textsc{Drift-Phoenix} calculated value. Therefore, the atmospheric pressure in the plots presented here act as a proxy for atmospheric height and in general doesn't reflect the true local pressure. For comparison, also plotted is the time taken to grow or remove a monolayer of material through neutral gas-phase surface chemistry using \textsc{Drift-Phoenix}: $\tau_{\rm mono}^{\rm DF}=10^{-10}/|\chi^{\rm net}|$, where the total net growth of the dust grain is $|\chi^{\rm net}|$, for the example atmospheres (Figure~\ref{dep_sputt_0}) given by $T_{\rm eff}=2400$~K, $\log{g}=5.0$ (cyan); $T_{\rm eff}=2400$~K, $\log{g}=3.0$ (magenta); $T_{\rm eff}=1500$~K, $\log{g}=5.0$ (green); and $T_{\rm eff}=1500$~K, $\log{g}=3.0$ (red). At high atmospheric pressures the numerical jitter in $\tau^{\rm DF}_{\rm mono}$ corresponds to transitions between growth rates ($\chi^{\rm net}>0$) and evaporation rates ($\chi^{\rm net}<0$). \label{dep_sputt_2}}
\end{figure}

Figure~\ref{dep_sputt_2} shows the monolayer timescale for plasma deposition/sputtering, $\tau_{\rm mono}$~(Equation~\ref{eq_mono}), as a function of atmospheric pressure for a range of electron temperatures and degrees of ionisation. For comparison, neutral gas-phase surface chemistry monolayer timescales are plotted for the considered \textsc{Drift-Phoenix} example atmospheres, $\tau_{\rm mono}^{\rm DF}$. At high atmospheric pressures it is more likely that dust evolution via plasma deposition and sputtering is the dominant process and $\tau_{\rm mono}<\tau_{\rm mono}^{\rm DF}$; for example, in the $\log{g}=5$ (or $\log{g}=3$) models, when $p_{\rm gas}\gtrsim10^{-3}$~bar ($\gtrsim10^{-5}$~bar), we are in the plasma regime if $f_{e}\gtrsim10^{-7}$ (as indicated by plasma laboratory devices, see \citet{fridman2008,diver2013}). At lower atmospheric pressures, a greater degree of ionisation is required to ensure that plasma processes dominate; for example, for the $\log{g}=5$ ($\log{g}=3$) models, when $p\lesssim10^{-5}$~bar ($\lesssim10^{-7}$~bar) a degree of ionisation $f_{e}\gtrsim10^{-6}$ is required. Generally, within the highest dust density cloud regions, $p_{\rm gas}\approx10^{-5}-1$~bar (Figure~\ref{dep_sputt_0}), plasma processes dominate if $f_{e}\gtrsim10^{-4}$. The monolayer timescales presented here are consistent with those reported from plasma laboratory experiments.  \cite{cao2002,cao2003} and \cite{matsoukas2004} report dust growth rates of the order of $0.1-1$~nm/min for $f_{e}\approx10^{-7}$ $n_{i}=10^{16}$~m$^{-3}$, $T_{e}=7$~eV, $T_{i}=600$~K and a chamber pressure of $\approx10^{-4}$~bar which corresponds to $\tau_{\rm mono}\approx10^{2}$~s reported here.

For fully ionised regions ($f_{e}=1$), the growth and evaporation of dust grains through neutral gas-phase surface chemistry cannot occur and we are in the plasma dominated regime; whereas, in a region where the degree of ionisation is insufficient for a plasma (i.e. $f_{e}\lesssim10^{-7}$), neutral gas-phase surface chemistry will be the dominant process.  For the interim degrees of ionisation ($10^{-7}\lesssim f_{e}<1$), there will be a transition from the gas-phase dominated regime ($\tau_{\rm mono}>\tau_{\rm mono}^{\rm DF}$) to the plasma dominated regime ($\tau_{\rm mono}^{\rm plasma}<\tau_{\rm mono}^{\rm gas}$).  As the degree of ionisation increases, so does the ion density, the ion flux at the grains surface and the effect of plasma deposition/sputtering. For a given degree of ionisation, greater electron temperatures (more energetic plasma electrons), increases the number of electrons that can reach the grain surface and this has the effect of increasing the magnitude of the floating potential, that is, the magnitude of the surface potential of the electron saturated grain. As a result, the flux of ions towards the grain increases, and hence the number of surface reactions that occur. Depending on the gas-phase species that is ionised, the formation of even a weakly-ionised plasma could disrupt the neutral gas-phase surface chemistry that leads to mantle growth and evaporation, since the participating neutral species could be ionised; for example, if Mg is ionised this could disrupt grain mantle growth through the chemical surface reactions that yield solid enstatite and forsterite, etc~\citep{helling2008c}.

\begin{figure}
\resizebox{\hsize}{!}{\includegraphics{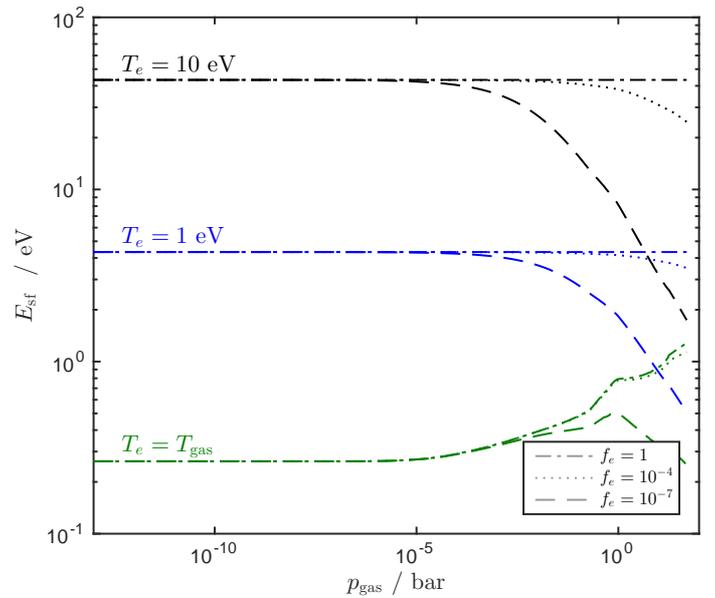}}
\caption{Energy of the ions at the surface of the dust grain, $E_{\rm sf}$, as a function of gas pressure.  The ionic energies, $E_{\rm sf}$, are given for electron temperatures $T_{e}=T_{\rm gas}$ (green), 1~eV (blue) and 10~eV (black); and, degrees of ionisation $f_{e}=1$ (dot-dash), $10^{-4}$ (dot) and $10^{-7}$ (dash). \label{dep_sputt_04}}
\end{figure}

The energy of the ions at the surface of the dust grain, $E_{\rm sf}$, after acceleration through the sheath, and its dependence on the degree of ionisation and electron temperature, is shown in Figure~\ref{dep_sputt_04}. For a fully ionised plasma ($f_{e}=1$) there are no ion neutral collisions ($p=0$) and so the ions acquire the maximum amount of energy. For lower degrees of ionisation, there are more neutrals and a greater probability of an ion-neutral collision, disrupting the acceleration of the ions and so lowering the energy they ultimately have at the grain surface.  Furthermore, the lower the ionic energy at the grain surface, the lower the sputtering yield, $Y_{\rm sp}$ and hence the effect of plasma sputtering (i.e. lower $\Gamma$).  As the degree of ionisation decreases, the electron number density decreases and hence the sheath length $d_{\rm sh}$ increases; moreover, the neutral number density increases leading to a smaller ion-neutral collision mean free path $\lambda_{\rm mfp,n}$. As a result, the ratio of the plasma sheath length and the mean free path $\delta_{\rm ni}=d_{\rm sh}/\lambda_{\rm mfp,n}$ increases and the plasma sheath becomes more collisional. 

\begin{figure*}
\centering
\includegraphics[width=17cm]{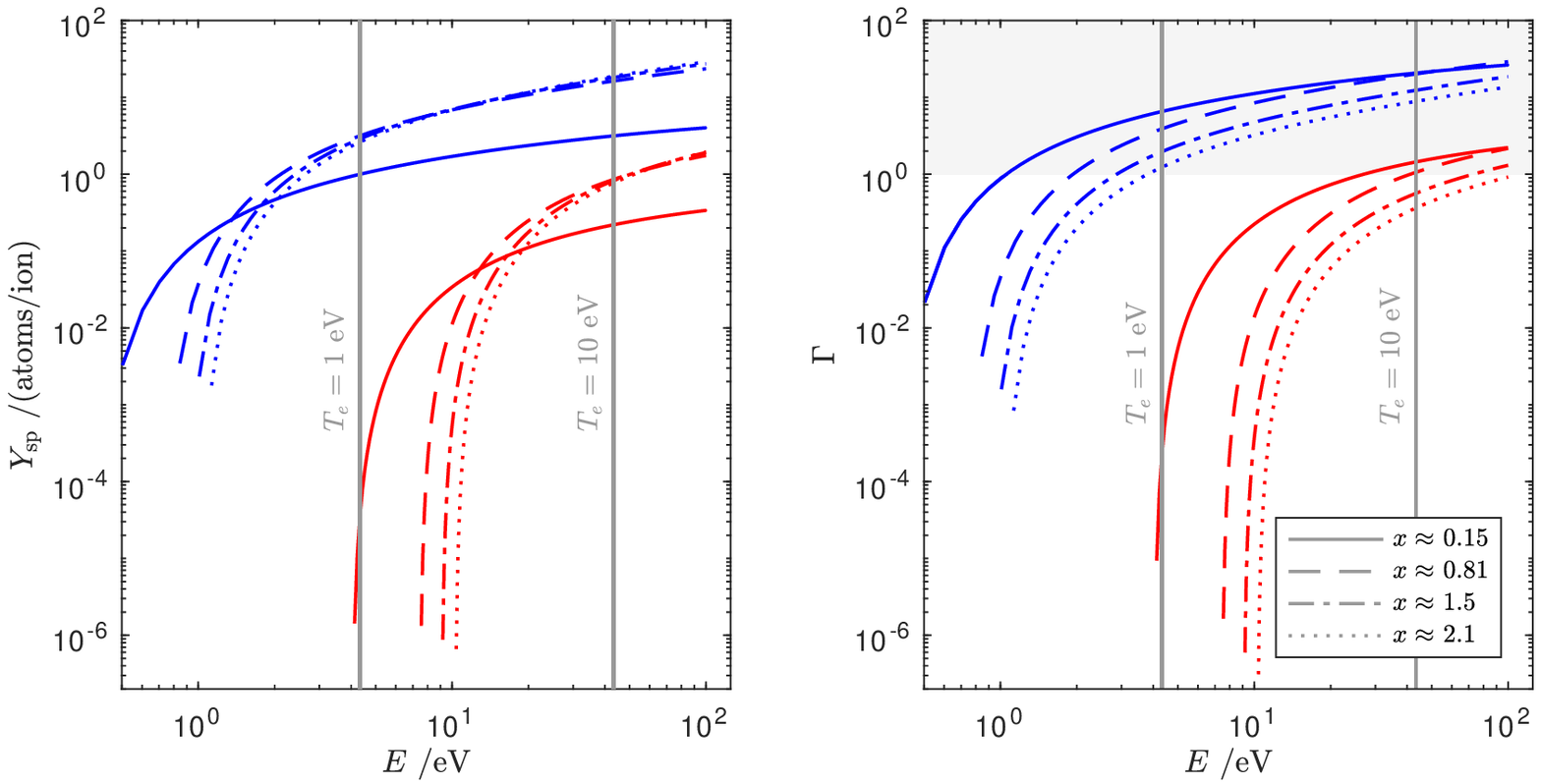}
\caption{Sputtering yield $Y_{\rm sp}$ (left) and $\Gamma$ (right), as a function of incident ion energy for surface binding energies of $E_{\rm b}=1$~eV (red) and $E_{\rm b}=0.1$~eV (blue); and, $x=m_{\rm i}/m_{\rm n}\approx 0.15$ (solid), $\approx 0.81$ (dash), $\approx 1.5$ (dot-dash) and $\approx 2.1$ (dot). Typically the target species range from light H atoms to heavier Mg$_{2}$SiO$_{4}$ molecules, to encapsulate this range the geometric mean of the potential target masses was calculated as $m_{\rm n}\approx26u\approx \mathcal{O}(10^{-26}~$kg$)$.  Therefore, for this constant value of $m_{\rm n}$ the range of $x$ plotted cover ion masses $m_{\rm i}=4.0026u$ (He) to $m_{\rm i}=55.845u$ (Fe), covering the expected range of ionic masses in the atmosphere. The regime $\Gamma>1$ (the sputtering regime) is signified by a grey shaded area. For comparison the energy of ions at the surface of the dust grain, $E_{\rm sf}$ (Figure~\ref{dep_sputt_04}), for $T_{e}=1$ eV and $T_{e}=10$ eV when $f_{e}=1$ (the degree of ionisation, $f_{e}=n_{e}/(n_{\rm n}+n_{e})$), are over-plotted (grey-solid).\label{sputt_plot_1}}
\end{figure*}

Figure~\ref{sputt_plot_1} shows the sputtering yield $Y_{\rm sp}$ (Equation~\ref{sputt_eqn}) and $\Gamma$ (Equation~\ref{big_g}) as a function of incident ion energy and surface binding energy. Typically the target species range from light H atoms to heavier Mg$_{2}$SiO$_{4}$ molecules, to encapsulate this range the geometric mean of the potential target masses was calculated as $m_{\rm n}\approx26u\approx \mathcal{O}(10^{-26}~$kg$)$.  Therefore, for this constant value of $m_{\rm n}$ the range of $x$ plotted cover ion masses $m_{\rm i}=4.0026u$ (He) to $m_{\rm i}=55.845u$ (Fe), covering the expected range of ionic masses in the atmosphere.  For a given target mass $m_{\rm n}$, increasing the incoming ionic mass $m_{\rm i}$ (and so $x$), increases the likelihood of liberating a target atom or molecule. This behaviour is due to the complex nature of the sputtering process described by Eq.~(\ref{sputt_eqn}) and has been experimentally verified (see Figure 10, \citet{tielens1994}).

In plasma regions characterised by $T_{e}\gtrsim1~eV$ and $f_{e}=1$, loosely bound dust with surface bond strengths of the order of $0.1$~eV (van der Waals bonds) will be heavily sputtered ($Y_{\rm sp}>1$) and the plasma region is in the sputtering regime, $\Gamma>1$.  In contrast, only in plasma regions with electron temperatures approaching $10$~eV, will the incoming ions be energetic enough to sputter target species with surface binding energies of the order of $1$~eV (such as covalent or ionic bonds). As a result, it is expected that crystalline grains with bond strengths of the order of $10$~eV will be largely immune to plasma sputtering ($\Gamma<1$).  For regions with lower degrees of ionisation, the sputtering yields will be reduced since the energisation of the incoming ions will be disrupted  due to ion-neutral collisions in the plasma sheath (see Figure~\ref{dep_sputt_2}).  Sputtering yields below unity ($Y_{\rm sp}<1$) signify that the incoming ions have insufficient energy to dislodge surface atoms and molecules; however, the energy of the incoming ion is deposited into the surface, exciting the surface layer of atoms and potentially making it easier for subsequent ions that follow to liberate an atom, even if their energy (on its own) is less than that required to surmount the energy barrier. Further to this, the electrostatic energy of the charged grains will reduce the surface binding energies of the surface neutrals, making it easier to sputter material than in the non-charged case. The binding energies of loose particle aggregates as well as crystalline dust grains that exhibit crystalline defects such as dislocations, will be reduced undermining their structural integrity.

\section{Discussion}
This paper presents a novel mechanism for dust evolution in sub-stellar atmospheres that is complementary to, but different from, standard chemical processes if ionisation is present. In contrast to previous studies, this paper considers how plasma collective effects self-consistently energise the plasma ions such that they can participate in plasma deposition and plasma sputtering. Dust growth and destruction via plasma deposition and sputtering is relevant to environments where energetic processes such as lightning, or strong flows near magnetic fields occur and is the natural outcome of incorporating macroscopic electrodynamics into atmospheric cloud evolution. However, in comparison to contemporary dust evolution through gas-phase chemistry, plasma processes such as these can be faster and have a significant impact on the evolution of dust clouds as well as the composition of the surrounding atmospheric environment.

Within the highest dust density cloud regions ($p_{\rm gas}\approx10^{-5}-1$~bar), plasma deposition and sputtering dominates over neutral gas-phase surface chemistry if $f_{e}\gtrsim10^{-4}$. We note that this also corresponds to the region where lightning is expected to occur in analogy to terrestrial lightning which occurs where the atmospheric pressure $\approx 0.1$~bar.  Loosely bound grains with surface binding energies of the order of $0.1-1$~eV are susceptible to destruction through plasma sputtering for feasible degrees of ionisation and electron temperatures; whereas, strong crystalline grains with binding energise of the order $10$~eV are resistant to sputtering.

In the deposition regime ($\Gamma<1$), the accretion of plasma ions onto the grain surface will deplete the local ambient atmosphere gas-plasma mixture and will alter the chemical composition of the dust grain.  This is dependent upon the majority species that is ionised in the atmospheric plasma region and may affect the resulting electromagnetic spectrum directly through the depletion of a particular species or indirectly, through the consequent gas-plasma chemistry due to the loss of a particular reactant. For example, thermal ionisation preferentially ionises Na, K, Ca, Mg and F~\citep{isabel2015}; Lightning ionises He, H$_{2}$, N$_{2}$, H$_{2}$O, CO$_{2}$~\citep{helling2013}; and AI will most likely ionise Fe, Mg, Na, H$_{2}$, CO, H$_{2}$O, N$_{2}$ and SiO~\citep{stark2013}. In general, neutral species that possess a sufficiently low first ionisation potential, whilst simultaneously being sufficiently abundant will produce the majority ionic species. However, metastables, such as He and H$_{2}$, can be ionised in two stages yielding higher abundances than expected based primarily on their first ionisation potential. Furthermore, dissociative attachment and dissociative recombination are additional processes that can affect the atmospheric abundances. 

In the plasma sputtering regime ($\Gamma>1$), the liberation of atoms will replenish the local ambient gas-plasma mixture, possibly counteracting the depletion through deposition.  Therefore, by virtue of plasma deposition and sputtering, gas-phase species could be transported from one atmospheric region to another via the transport of dust grains and enhance chemical mixing in different parts of the atmosphere, faster than chemical processes alone. This type of process could create spots in the atmosphere with an over- or under-density of a particular species relative to surrounding regions. As a result, through the alteration of the local particle size distribution or the local gas-plasma chemical composition, plasma deposition and plasma sputtering could be a source of atmospheric variability. This is consistent with \citep{buenzli2014}, where variations in cloud cover can occur at timescales of the order of 100s, (see Figures 7-10 in \citet{buenzli2014}) which is feasible by the plasma deposition and sputtering process over a wider range of pressure conditions than simple gas-phase chemical models. It is possible that plasma deposition and sputtering may be a contributing underlying physical mechanism that produces the observed inhomogeneity. 

The dusty plasma processes discussed here (and others e.g.~\citet{stark2015}) could also be relevant in other astrophysical dusty plasma environments such as protoplanetary disks, the interstellar medium and the circumstellar envelopes of AGB stars. 

\begin{acknowledgements}
The authors are grateful to the anonymous referee for constructive comments and suggestions that have improved this paper. CRS is grateful for funding from the Royal Society via grant number RG160840 and for support from the Division of Computing and Mathematics at Abertay University. DAD is grateful for funding from EPSRC via grant number EP/N018117/1. Thanks to Ch. Helling for the use of the \textsc{Drift-Phoenix} data used in this paper. \end{acknowledgements}
\bibliographystyle{aa}
\bibliography{coul_pap}
\end{document}